\documentclass[conference]{IEEEtran}
\usepackage[pdftex]{graphicx}
\usepackage[cmex10]{amsmath}
\usepackage{hyperref}
\usepackage{color}

\pdfoutput=1

\providecommand{\U}[1]{\protect\rule{.1in}{.1in}}
\newtheorem{theorem}{Theorem}

\newtheorem{remark}[theorem]{Remark}

\newenvironment{proof}[1][Proof]{\noindent\textbf{#1.} }{\ \rule{0.5em}{0.5em}}

\let\originalleft\left
\let\originalright\right
\renewcommand{\left}{\mathopen{}\mathclose\bgroup\originalleft}
\renewcommand{\right}{\aftergroup\egroup\originalright}

\begin{document}
%
\title{Unconstrained distillation capacities  
of a pure-loss bosonic broadcast channel}

 \author{
   \IEEEauthorblockN{
     Masahiro Takeoka\IEEEauthorrefmark{1},
     Kaushik P.~Seshadreesan\IEEEauthorrefmark{2}, and
     Mark M.~Wilde\IEEEauthorrefmark{3}}
   \IEEEauthorblockA{\it 
     \IEEEauthorrefmark{1}
    Quantum ICT Laboratory, 
    National Institute of Information and 
    Communications Technology, 
    Koganei, Tokyo 184-8795, Japan}
   \IEEEauthorblockA{\it 
     \IEEEauthorrefmark{2}
    Max-Planck-Institut f\"{u}r die Physik des Lichts, 
    91058 Erlangen, Germany}
   \IEEEauthorblockA{\it 
     \IEEEauthorrefmark{3}
    Hearne Institute for Theoretical Physics, 
    Department of Physics and Astronomy, \\
    Center for Computation and Technology, 
    Louisiana State University, 
    Baton Rouge, Louisiana 70803, USA\\
    }

 }

\maketitle

\begin{abstract}
Bosonic channels are important in practice as they form a simple model 
for free-space or fiber-optic communication.
Here we consider a single-sender multiple-receiver pure-loss bosonic  broadcast channel
and determine the unconstrained capacity region for the distillation of 
bipartite entanglement and secret key between the sender and each receiver, whenever they are allowed arbitrary
public classical communication. 
We show how the state merging protocol leads to achievable rates in this setting, giving an inner bound on the capacity region. 
We also determine an outer bound on the region and find that
the outer bound matches the inner bound in the infinite-energy limit, thereby establishing 
the unconstrained capacity region for such channels.  
Our result could provide a useful benchmark for implementing 
a broadcasting of entanglement and secret key through such channels. An important open question relevant to practice is to determine the capacity region in both this setting and the single-sender single-receiver case when there is an energy constraint on the transmitter.
\end{abstract}


%
\IEEEpeerreviewmaketitle

\section{Introduction}

\label{sec:intro}

Quantum key distribution (QKD) \cite{BB84,E91} and entanglement distillation
(ED) \cite{BBPSSW95} are two cornerstones of quantum communication.
QKD enables two or more cooperating parties to distill and share 
unconditionally secure random bit sequences, which could then be used 
for secure classical communication. 
ED, on the other hand, allows them to distill pure 
maximal entanglement from a quantum state shared via a noisy communication
channel, which could then be used to faithfully transfer quantum states
by using quantum teleportation \cite{BBCJPW93}.
In both protocols, the parties are allowed to perform
(in principle) an unlimited amount of local operations and 
classical communication (LOCC). 

Quantum communication technologies have matured tremendously in recent years.
In particular, QKD has been available commercially for a number of years and
has now expanded to inter-city networks \cite{SECOQC09,TOKYO_QKD}.
Also, efforts are currently underway to accomplish QKD in free space between the earth and satellites \cite{EGKSML15}.

Quantum communication, however, faces an important challenge. Like
most other quantum technologies, its performance is affected by
noise. Loss is the main source of error in typical
optical communication channels and severely limits the rates and distances at which
secret key or quantum entanglement can be distilled using the channel. All practical implementations of QKD to date
are known to exhibit a rate-loss tradeoff, in which the rate of secret
key extraction drops with increasing distance \cite{SBCDLP09}. In
the case of the standard optical-fiber communication channel, the
drop is exponential with increasing distance.

Some time after these limitations were observed, 
Refs.~\cite{TGW14Nat,TGW14IEEE} provided a mathematical proof, using the notion of squashed entanglement \cite{CW04}, that 
the tradeoff is indeed a fundamental limitation 
even with unconstrained input energy.
One of the main results of \cite{TGW14Nat,TGW14IEEE} is an upper bound on the two-way LOCC assisted quantum 
and secret key agreement capacity of a pure-loss bosonic channel, which is 
solely a function of the channel transmittance $\eta$. If the transmitter can consume only a finite amount of energy (as is in some practical cases), then tighter bounds are available \cite{TGW14Nat,TGW14IEEE}.
As a consequence,
 no yet-to-be-discovered protocol could ever surpass the limitations established in \cite{TGW14Nat,TGW14IEEE}.
 Ref.~\cite{Goodenough2015} extended the squashed entanglement technique to obtain upper bounds for a variety of phase-insensitive Gaussian channels.
Concurrently with \cite{Goodenough2015}, Ref.~\cite{PLOB15}  improved the infinite-energy bound from \cite{TGW14Nat,TGW14IEEE} and conclusively established 
the unconstrained capacity of the pure-loss bosonic channel
as $\mathcal{{C}}\left(\eta\right)=-\log_{2}\left(1-\eta\right)$. 
It is still an open question to determine the constrained capacity 
(i.e., when the transmitter is limited to consuming finite energy).

One of the long-term goals of quantum communication is to establish
a quantum internet \cite{K08}: a large collection of interconnected
quantum networks between multiple users that enables secure classical
communication and distributed quantum information processing. Apart
from point-to-point links, network architectures based on single-sender
multiple-receivers (modeled as broadcast channels) and vice versa
(multiple access channels) are also important in this context. 
Even though various network quantum communication scenarios 
 have been examined 
\cite{AS98,Winter01,GSE07,YHD08,YHD11,DHL10}, 
there has been limited work on the capacity of entanglement and secret key 
distillation assisted by unlimited LOCC. 
Only recently in \cite{STW15} 
were outer bounds on the achievable rates established for
multipartite secret key agreement and entanglement generation between
any subset of the users of a general single-sender $m$-receiver quantum
broadcast channel (QBC) (for any $m\geq1$) when assisted by unlimited
LOCC between all the users. The main idea was to employ
multipartite generalizations of the squashed entanglement
\cite{AHS08,YHHHOS09} and the methods of \cite{TGW14Nat,TGW14IEEE}.

In this paper, we consider a single-sender multiple-receiver pure-loss bosonic QBC 
and establish the unconstrained capacity region for the distillation of 
bipartite entanglement and secret key between the sender and each receiver 
assisted by unlimited LOCC. 
To prove the statement, we establish inner bounds on the achievable rate region 
by employing the quantum state merging protocol  \cite{HOW05,HOW07}. 
The converse part relies upon several tools. First, we utilize a teleportation
simulation argument originally introduced in \cite[Section~V]{BDSW96} and recently 
generalized in \cite{PLOB15}
to wider families of channels and continuous-variable systems. Next, it is known that the relative
entropy of entanglement is an upper bound on the distillable key of a
bipartite state \cite{HHHO05}, and the recent work in \cite{PLOB15} stated how
it is possible to combine the relative entropy
of entanglement upper bound with the teleportation simulation argument to arrive
at upper bounds on the secret-key agreement capacity of certain single-sender single-receiver
channels. 
We find that the outer bounds match the inner bounds in the infinite-energy limit, thereby establishing 
the unconstrained capacity region. An important open question is to determine the constrained capacity region, i.e., when only finite energy is available at the transmitter.

The paper is organized as follows. 
In Section~\ref{sec:background}, we describe a general LOCC-assisted distillation protocol 
for a QBC and the mathematical and physical model of the pure-loss bosonic QBC. 
The unconstrained capacity region is given in Section~\ref{sec:cap-region} along with a proof for the single-sender two-receiver case. 
Section~\ref{sec:concl} concludes the paper. 
The appendix generalizes the main theorem to the single-sender multiple-receiver case.

\section{LOCC-assisted distillation protocol and the channel model}

\label{sec:background}
In the main text, we consider a single-sender two-receiver QBC 
$\mathcal{N}_{A' \to BC}$ (Fig.~1(a)) 
and an ($n$, $E_{AB}$, $E_{AC}$, $K_{AB}$, $K_{AC}$, $\varepsilon$) 
protocol described as follows (the appendix considers the generalization to multiple receivers). 
The sender, Alice, prepares some quantum systems in an initial quantum state and successively sends 
some of these systems to the receivers, Bob and Charlie, 
by interleaving $n$ channel uses of the broadcast channel with rounds of LOCC. 
The goal of the protocol is to distill bipartite maximally 
entangled states $\Phi_{AB}$ and $\Phi_{AC}$ and private states 
$\gamma_{AB}$ and $\gamma_{AC}$ (equivalently secret keys \cite{HHHO09}). 
After each channel use, they can perform an arbitrary number of rounds of LOCC 
(in any direction with any number of parties). The quantities 
$E_{AB}$ and $E_{AC}$ denote entanglement rates (i.e.,  the logarithm of 
the Schmidt rank of $\Phi_{AB}$ and $\Phi_{AC}$, respectively, normalized by the number of channel uses)
and $K_{AB}$ and $K_{AC}$ are secret-key rates (i.e., the number of secret-key bits in $\gamma_{AB}$, 
and $\gamma_{AC}$, respectively, normalized by the number of channel uses). 
The protocol considered here is similar to the one 
described in \cite{STW15}, except that here we do not consider the other possible rates 
$E_{BC}$, $K_{BC}$, $E_{ABC}$, and $K_{ABC}$.

A rate tuple ($E_{AB}$, $E_{AC}$, $K_{AB}$, $K_{AC}$) is achievable 
if for all $\varepsilon \in (0,1)$ and sufficiently large $n$, 
there exists an ($n$, $E_{AB}$, $E_{AC}$, $K_{AB}$, $K_{AC}$, $\varepsilon$) 
protocol of the above form. The capacity region is  the closure of the set of all achievable rates.

In the following, we concentrate on a specific channel: 
a pure-loss bosonic QBC which we denote by 
$\mathcal{L}_{A' \to BC}$.
For this channel, the input state is split into three systems 
and one system is sent to each of Bob, Charlie, and the environment 
with transmittance $\eta_B$, $\eta_C$, and $1-\eta_B-\eta_C$, respectively,
where $\eta_B, \eta_C \in [0,1],$ $\eta_B + \eta_C \le 1$.  

Physically the signal splitting is modeled by a pair of 
two-input two-output beam splitters in which the signal is 
mixed with the vacuum state. 
For example, one can construct such a channel by a sequence of 
two beam splitters with transmittance $\eta_B + \eta_C$ and 
$\eta_B/(\eta_B+\eta_C)$, respectively, where 
the first beam splitter splits the signal and the environment, 
and the second one splits the signal into Bob's and Charlie's parts 
(see Fig.~1(b)). 
Mathematically this is characterized by the following input-output 
relation: 
\begin{align}
\label{eq:QBC_in-out1}
\hat{b} & =  \sqrt{\eta_B} \, \hat{a}' 
+ \sqrt{\frac{\eta_B (1-\eta_B-\eta_C)}{\eta_B+\eta_C}} \, \hat{f} 
+ \sqrt{\frac{\eta_C}{\eta_B+\eta_C}} \, \hat{g} ,
\\
\label{eq:QBC_in-out2}
\hat{c} & =  -\sqrt{\eta_C} \, \hat{a}' 
- \sqrt{\frac{\eta_C (1-\eta_B-\eta_C)}{\eta_B+\eta_C}} \, \hat{f} 
+ \sqrt{\frac{\eta_B}{\eta_B+\eta_C}} \, \hat{g} ,
\\
\label{eq:QBC_in-out3}
\hat{e} & =  -\sqrt{1-\eta_B-\eta_C} \, \hat{a}' 
+ \sqrt{\eta_B+\eta_C} \, \hat{f} ,
\end{align}
where $\hat{a}'$, $\hat{b}$, $\hat{c}$, $\hat{e}$ are 
annihilation operators for Alice's input, Bob's output, and Charlie's output 
modes, respectively, and $\hat{f}$ and $\hat{g}$ are annihilation operators 
for vacuum inputs from the environment.

Critical for our analysis is that the physical implementation of 
$\mathcal{L}_{A' \to BC}$ is not unique. 
One can model the same channel by a different concatenation
of two other beam splitters: for example, we could have
a first beam splitter split system $B$ from $C$ and $E$, 
and then a second one split $C$ and $E$. 
Obviously, it is also possible to split $C$ at the first beam splitter. 
These physical models are described in Fig.~1(c) and (d), respectively. 
In the next section, we will use these other physical models to 
explicitly calculate the rate regions.

\begin{figure}
\begin{center}
\includegraphics[width=83mm]{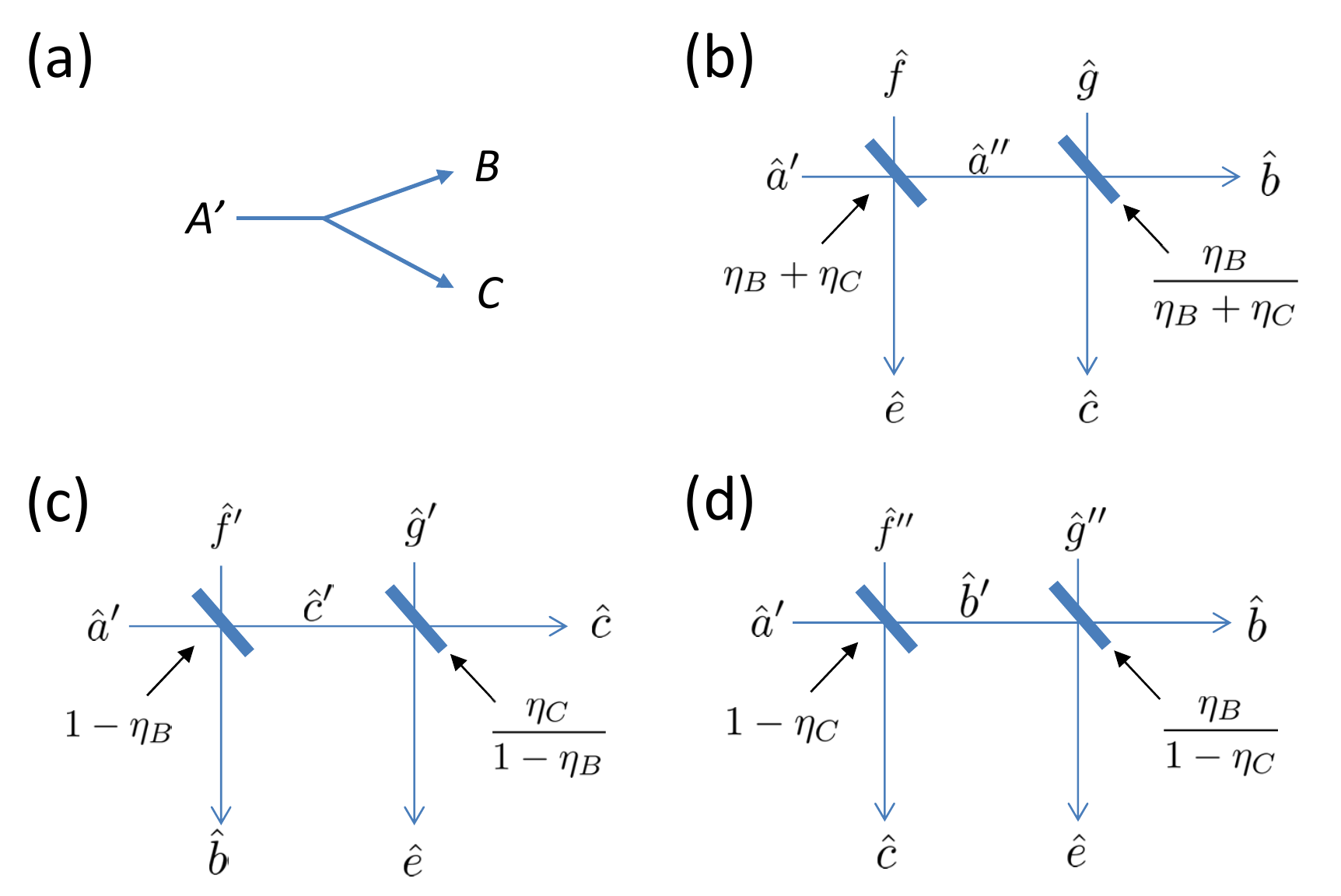}   %
\caption{\label{fig1}
(a) Single-sender two-receiver quantum broadcast channel. 
(b)-(d) Various physical implementations of the pure-loss 
bosonic broadcast channel with transmittances $\eta_B$ and $\eta_C$. 
}
\end{center}
\end{figure}

\section{Unconstrained capacity region}

\label{sec:cap-region}

Our main contribution  is the following theorem:
\begin{theorem}
\label{thm:capacity_region}
The LOCC-assisted, unconstrained capacity region 
of the pure-loss bosonic QBC $\mathcal{L}_{A' \to BC}$ 
is given by
\begin{align}
\label{eq:capacity_pureloss1}
E_{AB} + K_{AB}  \le  
\log_2 ( [1-\eta_C]/[1-\eta_B-\eta_C] ), 
\\
\label{eq:capacity_pureloss2}
E_{AC} + K_{AC}  \le  
\log_2 ( [1-\eta_B]/[1-\eta_B-\eta_C] ), 
\\
\label{eq:capacity_pureloss3}
E_{AB} + K_{AB} + E_{AC} + K_{AC}  \le 
-\log_2 ( 1-\eta_B-\eta_C). 
\end{align}
\end{theorem}

See the appendix for a generalization of this theorem to the multiple-receiver model from \cite{G08thesis}.

\subsection{Achievability part}

To achieve the rate region 
\eqref{eq:capacity_pureloss1}--\eqref{eq:capacity_pureloss3}, 
we consider a distillation protocol which employs quantum state merging. 
State merging was introduced in \cite{HOW05,HOW07} and provides
 an operational meaning for the conditional quantum entropy. 
For a state $\rho_{AB}$, its conditional quantum entropy is defined as 
$H(A|B)_\rho = H(AB)_\rho - H(B)_\rho$ where $H(AB)_\rho$ and $H(B)_\rho$ 
are the quantum entropies of $\rho_{AB}$ and its marginal $\rho_B$, 
respectively. 
For many copies of $\rho_{AB}$ shared between Alice and Bob, 
$H(A|B)_\rho$ is the optimal rate at which maximally entangled two-qubit states need
to be consumed to transfer Alice's systems to Bob's side via LOCC. 
If $H(A|B)_\rho$ is negative, the result is that after transferring Alice's 
systems, they can gain (i.e., distill) entanglement at rate $-H(A|B)_\rho$. 
State merging also yields a quantum analog of the Slepian-Wolf theorem 
in classical distributed compression problem and has been applied to the QBC in \cite{YHD11,DHL10}.

Here we consider the following alternative state merging based protocol. 
Alice first prepares $n$ copies of a two-mode squeezed vacuum (TMSV) 
\begin{equation}
\label{eq:TMSV}
|\Psi(N_S)\rangle_{AA'} = \sum_{m=0}^{\infty} \sqrt{\lambda_m(N_S)} 
|m\rangle_A |m\rangle_{A'},
\end{equation}
where $|m\rangle$ is an $m$-photon state, 
$
\lambda_m(N_S) = N_S^m/\left(N_S+1\right)^{m+1}, 
$
 and $N_S$ is the average photon number of the state per mode. 
She sends system $A'$ to Bob and Charlie through 
a pure-loss broadcast channel. 
After $n$ uses of the channel, Alice, Bob, and Charlie share 
$n$ copies of the state 
$\phi_{ABC} = \mathcal{L}_{A' \to BC} (|\Psi(N_S)\rangle \langle \Psi(N_S) | _{AA'})$.

Then by using $\phi_{ABC}^{\otimes n}$, 
they perform state merging to establish entanglement. 
More precisely, Bob and Charlie transfer their system back to Alice by LOCC (similar to reverse reconciliation in the point-to-point scenario). 
This could be done by applying the point-to-point state merging protocol successively 
\cite{HOW05,HOW07} or alternatively, by applying 
the multiparty simultaneous decoding state merging \cite{Dutil11}. 
Then we obtain the achievable rate region for $E_{AB}$ and $E_{AC}$ as 
\begin{align}
\label{eq:SM_LB_AB}
E_{AB} & \le  - H(B|AC)_\phi ,
\\
\label{eq:SM_LB_AC}
E_{AC} & \le  - H(C|AB)_\phi ,
\\
\label{eq:SM_LB_AB_AC}
E_{AB} + E_{AC} & \le  - H(BC|A)_\phi .
\end{align}
Since one ``ebit'' of entanglement can generate one private bit of secret key, 
the left-hand side of the above inequalities can be modified as 
$E_{AB} \to E_{AB} + K_{AB}$, $E_{AC} \to E_{AC} + K_{AC}$.

The right-hand side of these inequalities can be explicitly calculated. 
Recall that the marginal of the TMSV 
$\Psi_{A'}(N_S) = {\rm Tr}_A [ |\Psi(N_S)\rangle
\langle \Psi(N_S) |_{AA'} ]$ 
is a thermal state with mean photon number $N_S$. 
Its entropy is equal to $H(A')_{\Psi} = g(N_S)$, where 
$g(x) = (x + 1) \log_2 (x+1) - x \log_2 x$. 
Also a pure-loss channel with transmittance $\eta$ 
maps a thermal state to another thermal state 
with reduced average photon number. 
Let $U^{\mathcal{L}}_{A' \to BCE}$ be an isometric extension of 
$\mathcal{L}_{A' \to BC}$ and let 
\begin{equation}
|\phi\rangle_{ABCE} = U^{\mathcal{L}}_{A' \to BCE} |\Psi(N_S)\rangle
_{AA'} ,
\end{equation}
be a purification of $\phi_{ABC}$. 
By using $|\phi\rangle_{ABCE}$ and 
observing the above facts, we have
\begin{align}
\label{eq:-H(B|AC)}
-H(B|AC)_\phi & =  H(AC)_\phi - H(ABC)_\phi 
\nonumber\\ & =  
H(BE)_\phi - H(E)_\phi 
\nonumber\\ & =  
g((1-\eta_C)N_S) - g((1-\eta_B-\eta_C)N_S). 
\nonumber
\end{align}
In the limit as $N_S \to \infty$, this converges to 
$ \log_2 \left( \frac{1-\eta_C}{1-\eta_B-\eta_C} \right)$.
Similarly, we obtain 
\begin{align}
-H(C|AB)_\phi & \to 
\log_2 \left( \frac{1-\eta_B}{1-\eta_B-\eta_C} \right),
\\
-H(BC|A)_\phi & \to  
\log_2 \left( \frac{1}{1-\eta_B-\eta_C} \right) ,
\end{align}
in the limit of infinitely large $N_S$. 
Thus \eqref{eq:capacity_pureloss1}--\eqref{eq:capacity_pureloss3} 
are achievable when there is no energy constraint on the transmitter.

\subsection{Converse part}

As stated at the end of Section~\ref{sec:intro},
the converse relies upon several tools and is given in terms of the relative entropy of entanglement (REE) \cite{VP98}.  
The REE for a quantum state $\rho_{AB}$ is defined by 
\begin{equation}
\label{eq:REE}
E_R(A;B)_\rho = 
\inf_{\sigma_{AB}\in\textrm{SEP}}D\left(\rho_{AB}\Vert \sigma_{AB} \right)
\end{equation}
where $D(\rho\Vert\sigma) = {\rm Tr}[\rho ( \log_2 \rho - \log_2 \sigma )]$ 
is the quantum relative entropy and SEP denotes the set of  
separable states.
The original LOCC-assisted communication protocol can equivalently be
rewritten by using a teleportation simulation argument  
 \cite[Section~V]{BDSW96} suitably extended
to continuous-variable bosonic channels \cite{PLOB15}. Teleportation simulation in the case of a point-to-point channel 
can be understood as the possibility of reducing a sequence 
of adaptive protocols involving two-way LOCC into a sequence 
of non-adaptive protocols followed by a final LOCC.
For all `teleportation-simulable channels' (more precisely the channels arising
from the action of teleportation on a bipartite state) that allow for such a reduction, 
an upper bound on the entanglement and secret key agreement capacity 
can be given by the REE \cite{PLOB15}, because the REE is an upper bound on the distillable key
of any bipartite state \cite{HHHO05}. 
Furthermore, for pure-loss bosonic channels, one can use a concise formula for the REE  given in \cite{PLOB15}. 
With these techniques, an upper bound on the unconstrained capacity of 
a point-to-point pure-loss channel is equivalent 
to the REE of the state resulting from sending an infinite-energy TMSV through the channel, explicitly calculated to be equal to $-\log_2 (1-\eta)$.

Since the pure-loss bosonic QBC is covariant with respect to displacement operations (which are the teleportation corrections for bosonic channels \cite{prl1998braunstein}), it is teleportation-simulable. 
Then the original broadcasting protocol described in the previous section 
can be replaced by the distillation of $n$ copies of 
$\phi_{ABC} = \mathcal{N}_{A' \to BC} (|\Psi(N_S)\rangle
\langle \Psi(N_S)|_{AA'})$ 
via the final LOCC. 
This LOCC distills entanglement and secret key; i.e., 
it generates a state $\omega_{ABC}$ which is $\varepsilon$-close to 
$\tilde{\Phi}_{ABC}$:  
\begin{equation}
\Vert \omega_{ABC} - \tilde{\Phi}_{ABC}\Vert_1 \le \varepsilon
\end{equation}
with 
\begin{equation}
\label{eq:ideal_state}
\tilde{\Phi}_{ABC} = 
\Phi_{A_1 B_1}^{\otimes n E_{AB}} \otimes 
\Phi_{A_2 C_1}^{\otimes n E_{AC}} \otimes 
\gamma_{A_3 B_2}^{\otimes n K_{AB}} \otimes 
\gamma_{A_4 C_2}^{\otimes n K_{AC}} ,
\end{equation}
where $A_i$, $B_i$, and $C_i$ are subsystems of 
$A$, $B$, and $C$, respectively. 
Then by using several well known properties of REE 
(monotonicity under LOCC, continuity, and subadditivity for product states), we find that 
\begin{align}
n(E_{AB}+K_{AB}) & \le  E_R(B;AC)_{\tilde{\Phi}} 
\nonumber\\ & \le  
E_R(B;AC)_{\omega} + f(n, \varepsilon) 
\nonumber\\ & \le  
n E_R(B;AC)_{\phi} + f(n, \varepsilon) ,
\end{align}
where $f(n,\varepsilon)$ is a function such that 
$ \lim_{\varepsilon \to 0,n \to \infty} \frac{1}{n} f(n,\varepsilon)=0$. 
Similar bounds can be obtained in terms of $E_R(C;AB)_\phi$ and $E_R(A;BC)_\phi$,
leading to the following outer bound for the capacity region: 
\begin{align}
\label{eq:REE_UB_AB}
E_{AB} + K_{AB} & \le  E_R(B;AC)_\phi
\\
E_{AC} + K_{AC} & \le  E_R(C;AB)_\phi ,
\\
E_{AB} + K_{AB} + E_{AC} + K_{AC} & \le  E_R(A;BC)_\phi .
\end{align}

To calculate the right-hand side of each inequality, we use a calculation from \cite{PLOB15}: 
 for a point-to-point pure-loss bosonic channel 
with transmittance $\eta$, which we denote by $\mathcal{L}_{A' \to B}^\eta$, 
the REE of $\mathcal{L}_{A' \to B}^\eta (|\Psi({N_S})\rangle \langle \Psi(N_S)|_{AA'})$ 
with $N_S \to \infty$ is given by $-\log_2 (1-\eta)$.

For $E_R(A;BC)_\phi$, 
consider the physical implementation of the channel in Fig.~1(b) 
and let $\phi'_{AA''}$ be the state such that only the first beam splitter 
is applied. 
The second beam splitter is a local unitary  
in the sense that it operates on $B$ and $C$ 
whereas our partition is now between $A$ and $BC$. 
Thus it does not change the REE. 
Then we have $E_R(A;A'')_{\phi'} = E_R(A;BC)_\phi$. 
Also, since $\phi'_{AA''}$ is a  TMSV followed by 
a pure-loss channel with transmittance 
$\eta_B + \eta_C$, we get 
\begin{align}
\label{eq:REE_pureloss_1}
\lim_{N_S \to \infty}  E_R(A;BC)_\phi & = \lim_{N_S \to \infty}  E_R(A;A'')_{\phi'}
\nonumber\\ & = 
-\log_2 ( 1-\eta_B-\eta_C ) .
\end{align}

To calculate $E_R(C;AB)_\phi$, we employ the physical implementation of 
the channel in Fig.~1(c) in which the first beam splitter 
with transmittance $1-\eta_B$ separates $B$ from the others. 
This beam splitter is followed 
by the second beam splitter with transmittance $\eta_C/(1-\eta_B)$ 
which separates $C$ and the environment $E$.

Let $\phi'_{ABC'}$ be the TMSV in which only the first beam splitter 
is applied. 
Observe that it is a pure state and its marginal $\phi'_{C'}$ is 
a thermal state with average photon number $(1-\eta_B)N_S$. 
Combining these two observations, we can conclude that 
the state has the following Schmidt decomposition: 
\begin{equation}
\label{eq:Schmidt_decomposition_ABC'}
|\phi''\rangle_{ABC'} = \sum_{m=0}^\infty \sqrt{\lambda_m((1-\eta_B)N_S)} 
|\varphi_m\rangle_{AB} |m\rangle_{C'} ,
\end{equation}
where $\{|\varphi_m\rangle_{AB}\}_m$ is some orthonormal basis. 
Since $\{|\varphi_m\rangle_{AB}\}$ is an orthonormal set, there exists a local unitary 
operation acting on systems $A$ and $B$ such that 
\begin{equation}
\label{eq:local_unitary}
U_{AB}: \, |\varphi_m\rangle_{AB} \to |m\rangle_A |{\rm aux}\rangle_B ,
\end{equation}
where $|{\rm aux}\rangle$ is some constant auxiliary state. Then we have 
\begin{align}
\label{eq:local_unitary_operation}
U_{AB} |\phi''\rangle_{ABC'} & =  |{\rm aux}\rangle_B 
|\Psi((1-\eta_B)N_S)\rangle_{AC'} 
\nonumber\\ & \equiv  
|\phi'''\rangle_{ABC'}. 
\end{align}

Let $\tilde{\phi}_{ABC} 
= \mathcal{L}^{\bar{\eta}}_{C' \to C} (|\phi'''\rangle
\langle\phi'''|_{ABC'})$
where $\bar{\eta} = \eta_C/(1-\eta_B)$. 
Note that 
$\phi_{ABC} = \mathcal{L}^{\bar{\eta}}_{C' \to C} (|\phi''\rangle\langle \phi'' |_{ABC'})$. 
Since the local unitary operation $U_{AB}$ does not change the REE between 
$AB$ and $C$, we have $E_R(C;AB)_{\tilde{\phi}} = E_R(C;AB)_{\phi}$. 
Moreover, $E_R(C;AB)_{\tilde{\phi}}$ is the REE for the TMSV 
with $(1-\eta_B)N_S$ followed by $\mathcal{L}^{\bar{\eta}}_{C' \to C}$. 
Then by using the result in \cite{PLOB15},  we find  
\begin{align}
\label{eq:E_R(AB;C)_phi}
\lim_{N_S \to \infty} E_R(C;AB)_{\phi} & = 
\lim_{N_S \to \infty}
E_R(C;AB)_{\tilde{\phi}} 
\nonumber\\ & = 
-\log_2 ( 1-\bar{\eta}) 
\nonumber\\ & =  
\log_2 ( [1-\eta_B]/[1-\eta_B-\eta_C] ). 
\end{align}
Similarly, with the physical implementation picture in Fig.~1(d), we obtain 
\begin{equation}
\label{eq:E_R(AC;B)_phi}
\lim_{N_S \to \infty} E_R(B;AC)_{\phi} = 
\log_2 ( [1-\eta_C]/[1-\eta_B-\eta_C] ),  
\end{equation}
which completes the proof of the converse part. 
Figure 2 illustrates an example of the unconstrained capacity region 
given in \eqref{eq:capacity_pureloss1}--\eqref{eq:capacity_pureloss3}.

\begin{figure}
\begin{center}
\includegraphics[width=50mm]{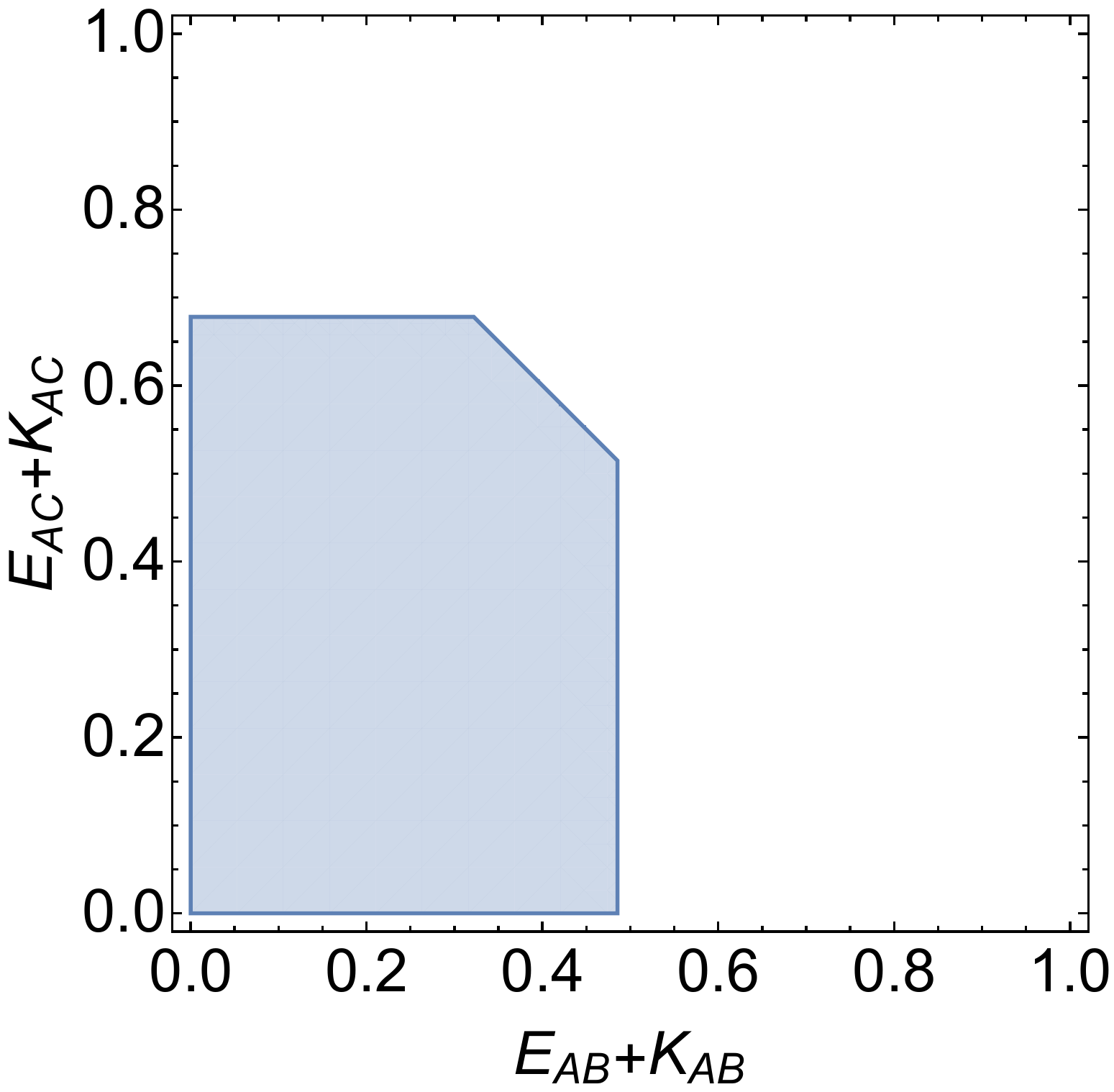}   %
\caption{\label{fig2}
LOCC-assisted capacity region 
given by
\eqref{eq:capacity_pureloss1}--\eqref{eq:capacity_pureloss3}
for the pure-loss bosonic broadcast channel, where
$(\eta_B,\, \eta_C) = (0.2, \, 0.3)$. 
}
\end{center}
\end{figure}

\section{Conclusion}

\label{sec:concl}
We have established the unconstrained capacity region of a pure-loss bosonic broadcast channel for 
 LOCC-assisted entanglement and secret key distillation. 
This result is proved by using the quantum state merging protocol (for the inner bound) 
and the relative entropy of entanglement (for the outer  bound), and it 
 could provide a useful benchmark for the
 broadcasting of entanglement and secret key through such channels.

There are some interesting problems left open. 
First, we consider the scenario to share entanglement and a secret key 
between the sender and each receiver, but we think it is interesting to include 
other possibilities, i.e., $E_{BC}$ and $K_{BC}$, and even 
 tripartite entanglement, such as GHZ states, and tripartite 
common secret key. 
Such a scenario was discussed in \cite{STW15}, where 
an outer bound was established by making use of the squashed entanglement. 
However, these bounds might be further improved, and in addition, it is open to determine 
how to construct a protocol achieving tight inner bounds. 

Second, one might attempt to generalize our scenario to other channels. 
A consequence of the teleportation simulation argument from \cite[Section~V]{BDSW96}
is that the REE upper bound can be applied to all teleportation-simulable channels,
as shown in \cite{PLOB15}. 
Thus the approach could work for all broadcast channels for which this is true. 
Finally, we think it is a pressing open question to determine 
the constrained capacity region (i.e., with a finite-energy constraint). 
Since the REE bound using the teleportation reduction technique requires 
infinite energy to realize 
an ideal teleportation, one needs an alternative approach here, 
such as the one given in \cite{STW15} in order to obtain tighter outer bounds in some cases.

We acknowledge helpful discussions with Sam Braunstein, Saikat Guha, 
Michal Horodecki, Stefano Pirandola, and John Smolin.
This research was supported by the Open Partnership Joint Projects of
JSPS Bilateral Joint Research Projects and the ImPACT Program of Council 
for Science, Technology and Innovation, Japan. MMW is grateful to NICT for hosting and supporting him for a research visit during December 2015, and he acknowledges support from NSF Grant No.~CCF-1350397.
KPS thanks the Max Planck Society for funding.

\bibliographystyle{IEEEtran}

\bibliography{Ref}

\begin{thebibliography}{10}
\providecommand{\url}[1]{#1}
\csname url@samestyle\endcsname
\providecommand{\newblock}{\relax}
\providecommand{\bibinfo}[2]{#2}
\providecommand{\BIBentrySTDinterwordspacing}{\spaceskip=0pt\relax}
\providecommand{\BIBentryALTinterwordstretchfactor}{4}
\providecommand{\BIBentryALTinterwordspacing}{\spaceskip=\fontdimen2\font plus
\BIBentryALTinterwordstretchfactor\fontdimen3\font minus
  \fontdimen4\font\relax}
\providecommand{\BIBforeignlanguage}[2]{{%
\expandafter\ifx\csname l@#1\endcsname\relax
\typeout{** WARNING: IEEEtran.bst: No hyphenation pattern has been}%
\typeout{** loaded for the language `#1'. Using the pattern for}%
\typeout{** the default language instead.}%
\else
\language=\csname l@#1\endcsname
\fi
#2}}
\providecommand{\BIBdecl}{\relax}
\BIBdecl

\bibitem{BB84}
C.~H. Bennett and G.~Brassard, ``Quantum cryptography: public key distribution
  and coin tossing,'' \emph{Proceedings of the IEEE International Conference on
  Computers, Systems, and Signal Processing}, p. 175, 1984.

\bibitem{E91}
A.~K. Ekert, ``Quantum cryptography based on {Bell}'s theorem,'' \emph{Physical
  Review Letters}, vol.~67, pp. 661--663, 1991.

\bibitem{BBPSSW95}
C.~H. Bennett, G.~Brassard, S.~Popescu, B.~Schumacher, J.~A. Smolin, and W.~K.
  Wootters, ``Purification of noisy entanglement and faithful teleportation via
  noisy channels,'' \emph{Physical Review Letters}, vol.~76, pp. 722--725,
  January 1996.

\bibitem{BBCJPW93}
C.~H. Bennett, G.~Brassard, C.~Cr\'epeau, R.~Jozsa, A.~Peres, and W.~K.
  Wootters, ``Teleporting an unknown quantum state via dual classical and
  {Einstein-Podolsky-Rosen} channels,'' \emph{Physical Review Letters},
  vol.~70, no.~13, pp. 1895--1899, March 1993.

\bibitem{SECOQC09}
M.~Peev, C.~Pacher, and R.~All{\'{e}}aume, ``{The SECOQC quantum key
  distribution network in Vienna},'' \emph{New Journal of Physics}, vol.~11, p.
  075001, 2009.

\bibitem{TOKYO_QKD}
{M. Sasaki et al.}, ``Field test of quantum key distribution in the {Tokyo QKD}
  network,'' \emph{Optics Express}, vol.~19, no.~11, pp. 10\,387--10\,409,
  2011, arXiv:quant-ph/1103.3566.

\bibitem{EGKSML15}
{D. Elser et al.}, ``{S}atellite {Q}uantum {C}ommunication via the {A}lphasat
  {L}aser {C}ommunication {T}erminal,'' 2015, arXiv:1510.04507v1.

\bibitem{SBCDLP09}
V.~Scarani, H.~Bechmann-Pasquinucci, N.~J. Cerf, M.~Du\v{s}ek,
  N.~L\"{u}tkenhaus, and M.~Peev, ``The security of practical quantum key
  distribution,'' \emph{Review of Modern Physics}, vol.~81, pp. 1301--1350,
  2009, arXiv:0802.4155.

\bibitem{TGW14Nat}
M.~Takeoka, S.~Guha, and M.~M. Wilde, ``Fundamental rate-loss tradeoff for
  optical quantum key distribution,'' \emph{Nature Communications}, vol.~5, p.
  5235, October 2014.

\bibitem{TGW14IEEE}
------, ``The squashed entanglement of a quantum channel,'' \emph{IEEE
  Transactions on Information Theory}, vol.~60, no.~8, pp. 4987--4998, August
  2014, arXiv:1310.0129.

\bibitem{CW04}
M.~Christandl and A.~Winter, ````{Squashed} entanglement'': An additive
  entanglement measure,'' \emph{Journal of Mathematical Physics}, vol.~45,
  no.~3, pp. 829--840, March 2004, arXiv:quant-ph/0308088.

\bibitem{Goodenough2015}
K.~Goodenough, D.~Elkouss, and S.~Wehner, ``Assessing the performance of
  quantum repeaters for all phase-insensitive {Gaussian} bosonic channels,''
  Nov. 2015, arXiv:1511.08710.

\bibitem{PLOB15}
S.~Pirandola, R.~Laurenza, C.~Ottaviani, and L.~Banchi, ``Fundamental limits of
  repeaterless quantum communications,'' 2015, arXiv:1510.08863.

\bibitem{K08}
H.~J. Kimble, ``The quantum internet,'' \emph{Nature}, vol. 453, no. 7198, pp.
  1023--1030, Jun. 2008.

\bibitem{AS98}
A.~E. Allahverdyan and D.~B. Saakian, ``{The broadcast quantum channel for
  classical information transmission},'' pp. 1--8, 1998.

\bibitem{Winter01}
A.~Winter, ``{The capacity of the quantum multiple-access channel},''
  \emph{IEEE Transactions on Information Theory}, vol.~47, no.~7, pp.
  3059--3065, 2001.

\bibitem{GSE07}
S.~Guha, J.~Shapiro, and B.~Erkmen, ``{Classical capacity of bosonic broadcast
  communication and a minimum output entropy conjecture},'' \emph{Physical
  Review A}, vol.~76, no.~3, p. 032303, September 2007.

\bibitem{YHD08}
J.~Yard, P.~Hayden, and I.~Devetak, ``{Capacity theorems for quantum
  multiple-access channels: classical-quantum and quantum-quantum capacity
  regions},'' \emph{IEEE Transactions on Information Theory}, vol.~54, no.~7,
  pp. 3091--3113, 2008.

\bibitem{YHD11}
------, ``Quantum broadcast channels,'' \emph{IEEE Transactions on Information
  Theory}, vol.~57, no.~10, pp. 7147--7162, October 2011.

\bibitem{DHL10}
F.~Dupuis, P.~Hayden, and K.~Li, ``A father protocol for quantum broadcast
  channels,'' \emph{IEEE Transactions on Information Theory}, vol.~56, no.~6,
  pp. 2946--2956, June 2010.

\bibitem{STW15}
K.~P. Seshadreesan, M.~Takeoka, and M.~M. Wilde, ``Bounds on entanglement
  distillation and secret key agreement for quantum broadcast channels,''
  \emph{Accepted for publication in IEEE Transactions on Information Theory},
  2015, arXiv:1503.08139.

\bibitem{AHS08}
D.~Avis, P.~Hayden, and I.~Savov, ``Distributed compression and multiparty
  squashed entanglement,'' \emph{Journal of Physics A: Mathematical and
  Theoretical}, vol.~41, no.~11, p. 115301, March 2008, arXiv:0707.2792.

\bibitem{YHHHOS09}
D.~Yang, K.~Horodecki, M.~Horodecki, P.~Horodecki, J.~Oppenheim, and W.~Song,
  ``Squashed entanglement for multipartite states and entanglement measures
  based on the mixed convex roof,'' \emph{IEEE Transactions on Information
  Theory}, vol.~55, no.~7, pp. 3375--3387, July 2009.

\bibitem{HOW05}
M.~Horodecki, J.~Oppenheim, and A.~Winter, ``{Partial quantum information.}''
  \emph{Nature}, vol. 436, no. 7051, pp. 673--6, August 2005.

\bibitem{HOW07}
------, ``{Quantum state merging and negative information},''
  \emph{Communications in Mathematical Physics}, vol. 136, pp. 107--136, 2007.

\bibitem{BDSW96}
C.~H. Bennett, D.~P. DiVincenzo, J.~A. Smolin, and W.~K. Wootters,
  ``{Mixed-state entanglement and quantum error correction},'' \emph{Physical
  Review A}, vol.~54, no.~5, p. 3824, 1996.

\bibitem{HHHO05}
K.~Horodecki, M.~Horodecki, P.~Horodecki, and J.~Oppenheim, ``Secure key from
  bound entanglement,'' \emph{Physical Review Letters}, vol.~94, no.~16, p.
  160502, April 2005, arXiv:quant-ph/0309110.

\bibitem{HHHO09}
------, ``General paradigm for distilling classical key from quantum states,''
  \emph{IEEE Transactions on Information Theory}, vol.~55, no.~4, pp.
  1898--1929, April 2009, arXiv:quant-ph/0506189.

\bibitem{G08thesis}
S.~Guha, ``Multiple-user quantum information theory for optical communication
  channels,'' Ph.D. dissertation, Massachusetts Institute of Technology, June
  2008.

\bibitem{Dutil11}
N.~Dutil, ``Multiparty quantum protocols for assisted entanglement
  distillation,'' \emph{arXiv preprint arXiv:1105.4657}, no. May, 2011.

\bibitem{VP98}
V.~Vedral and M.~B. Plenio, ``Entanglement measures and purification
  procedures,'' \emph{Physical Review A}, vol.~57, no.~3, pp. 1619--1633, March
  1998, arXiv:quant-ph/9707035.

\bibitem{prl1998braunstein}
S.~L. Braunstein and H.~J. Kimble, ``Teleportation of continuous quantum
  variables,'' \emph{Physical Review Letters}, vol.~80, pp. 869--872, 1998.

\end{thebibliography}

\pagebreak

\section*{Appendix: the 1-to-$m$ broadcast channel}

In this appendix, we generalize Theorem \ref{thm:capacity_region} 
to the 1-to-$m$ pure-loss broadcast channel for arbitrary positive 
integer $m$. 

Consider the pure-loss broadcast channel 
$\mathcal{L}_{A' \to B_1 \cdots  B_m}$ characterized by 
a set of transmittances $\{\eta_{B_1} , \cdots , \eta_{B_m} \}$ with 
$\sum_{i=1}^{m} \eta_{B_i} \le 1$ \cite{G08thesis}. 
Let $\mathcal{B} = \{B_1, \cdots ,B_m \}$, 
$\mathcal{T} \subseteq \mathcal{B}$, 
and $\overline{\mathcal{T}}$ be a complement of set $\mathcal{T}$. 
Then we have the following theorem:

\bigskip

\begin{theorem}
\label{thm:capacity_region_m-receiver}
The LOCC-assisted unconstrained capacity region of the pure-loss 
bosonic QBC $\mathcal{L}_{A' \to B_1 \cdots B_m}$ is given by 
\begin{equation}
\label{eq:capacity_pureloss_m-receiver}
\sum_{B_i \in \mathcal{T}} E_{AB_i} + K_{AB_i} \le 
\log_2 \left( 
\frac{1-\eta_{\overline{\mathcal{T}}}}{
1-\eta_{\mathcal{B}}} 
\right) ,
\end{equation}
for all non-empty $\mathcal{T}$, 
where $\eta_\mathcal{B} = \sum_{i=1}^m \eta_{B_i}$ and 
$\eta_{\overline{\mathcal{T}}} 
= \sum_{B_i \in \overline{\mathcal{T}}} \eta_{B_i}$. 
\end{theorem}
\bigskip

\begin{proof}
The strategy of the proof is quite similar to that of 
Theorem \ref{thm:capacity_region}. 
For the achievability, one can apply the point-to-point state merging 
protocol successively, which leads to the following achievable rate region:
\begin{equation}
\label{eq:SM_LB_m-receiver}
\sum_{B_i \in \mathcal{T}} E_{AB_i} \le 
- H(\mathcal{T}|A \overline{\mathcal{T}})_\phi ,
\end{equation}
where 
\begin{equation}
\label{eq:phi_m}
\phi_{AB_1 \cdots B_m} = 
\mathcal{L}_{A' \to B_1 \cdots B_m} 
(|\Psi(N_S)\rangle\langle\Psi(N_S)|_{AA'}) .
\end{equation}
The right-hand side of (\ref{eq:SM_LB_m-receiver}) is 
calculated to be  
\begin{align}
\label{eq:-H(T|ATbar)}
-H(\mathcal{T}|A \overline{\mathcal{T}})_\phi & =  
H(A \overline{\mathcal{T}})_\phi - 
H(A \mathcal{T} \overline{\mathcal{T}})_\phi 
\nonumber\\ & =  
H(\mathcal{T}E)_\phi - H(E)_\phi 
\nonumber\\ & =  
g((1-\eta_{\overline{\mathcal{T}}})N_S) - 
g((1-\eta_{\mathcal{B}})N_S) ,
\nonumber
\end{align}
and taking $N_S \to \infty$, we get 
$\log_2 ([1-\eta_{\overline{\mathcal{T}}}]/[1-\eta_{\mathcal{B}}])$. 
Since one ebit of entanglement can generate one private bit of key, 
we can replace $E_{AB_i}$ with $E_{AB_i}+K_{AB_i}$ which completes 
the achievability part. 
\begin{remark}
Since the above rate region reflects the last gain/consumption of 
entanglement after the sequential operation of the point-to-point 
state mergings, it could be possible that the protocol is `catalytic,' 
meaning that entanglement is consumed at some state merging 
which is compensated by the following other state mergings. 
However, this does not happen in our case. 
We can check it by the following simple observation. 
Gain/consumption of entanglement at 
any state merging is given by $-H(\mathcal{S}_1|A\mathcal{S}_2)_\phi$ 
where $\mathcal{S}_1$ is some nonempty subset of $\mathcal{B}$ 
and $\mathcal{S}_2$ is other subset (possibly empty) of $\mathcal{B}$ 
satisfying $\mathcal{S}_1 \cap \mathcal{S}_2 = \emptyset$.
Since $-H(\mathcal{S}_1|A\mathcal{S}_2)_\phi 
= H(\overline{\mathcal{S}_2})_\phi 
- H(\overline{\mathcal{S}_1\mathcal{S}_2})_\phi$ and $\mathcal{S}_1$ 
is non-empty, this quantity always has a positive value meaning 
that entanglement is generated at all steps of the whole protocol. 
\end{remark}

For the converse, we need to configure the beam splitter network of 
the QBC properly. Note that the channel has $m+1$ transmittances 
$\eta_{B_1}, \cdots , \eta_{B_m}$, and 
$\eta_{E} \equiv 1-\sum_{i} \eta_{B_i}$. 
We can order these transmittances in some sequence and label it as 
$\eta_1, \eta_2, \cdots , \eta_m, \eta_{m+1}$.
Then for any ordering, we can describe the channel 
by a sequence of $m$ beam splitters where the $j$-th beam splitter's 
transmittance is given by 
\begin{equation}
\label{eq:eta_i}
\tilde{\eta}_j = \frac{1-\sum_{k=1}^{j} \eta_k}{
1-\sum_{l=1}^{j-1} \eta_l} .
\end{equation}
Now, for each given $\mathcal{T}$ involving $t$ parties, 
consider the following specific ordering. 
For $\eta_i$ with $1 \le i \le t$, 
assign $\eta_{B_j}$ with $B_j \in \overline{\mathcal{T}}$, 
$\eta_{t+1} = \eta_E$, and for $\eta_i$ with $i > t+1$, assign 
$\eta_{B_j}$ with $B_j \in \mathcal{T}$.
Then the transmittance of the $t+1$ beam splitter is 
$\eta_{\mathcal{T}}/(1-\eta_{\overline{\mathcal{T}}})$ 
where $\eta_{\mathcal{T}} = \sum_{B_i \in \mathcal{T}} \eta_{B_i}$ 
(see Fig.~3). 
From the main text, we already know that the REE 
for given partition $\mathcal{T}$ and $A\overline{\mathcal{T}}$ 
is simply characterized by this transmittance. 
As a consequence, the REE bound for $N_S \to \infty$ turns out to be 
\begin{align}
\label{eq:REE_m-receiver}
\sum_{B_i \in \mathcal{T}} (E_{AB_i} + K_{AB_i}) & \le 
E_R (\mathcal{T} ; A \overline{\mathcal{T}})_\phi
\nonumber \\ & =  
\log_2 \left( \frac{1}{
1-\frac{\eta_{\mathcal{T}}}{1-\eta_{\overline{\mathcal{T}}}}}
\right)
\nonumber \\ & =  
\log_2 \left( \frac{1-\eta_{\overline{\mathcal{T}}}}{
1-\eta_{\mathcal{T}} - \eta_{\overline{\mathcal{T}}}} \right) ,
\nonumber \\ & =  
\log_2 \left( \frac{1-\eta_{\overline{\mathcal{T}}}}{
1-\eta_{\mathcal{B}}} \right) ,
\end{align}
where 
$\eta_{\mathcal{B}} = \eta_{\mathcal{T}} + \eta_{\overline{\mathcal{T}}}$. 
\begin{figure}
\begin{center}
\includegraphics[width=83mm]{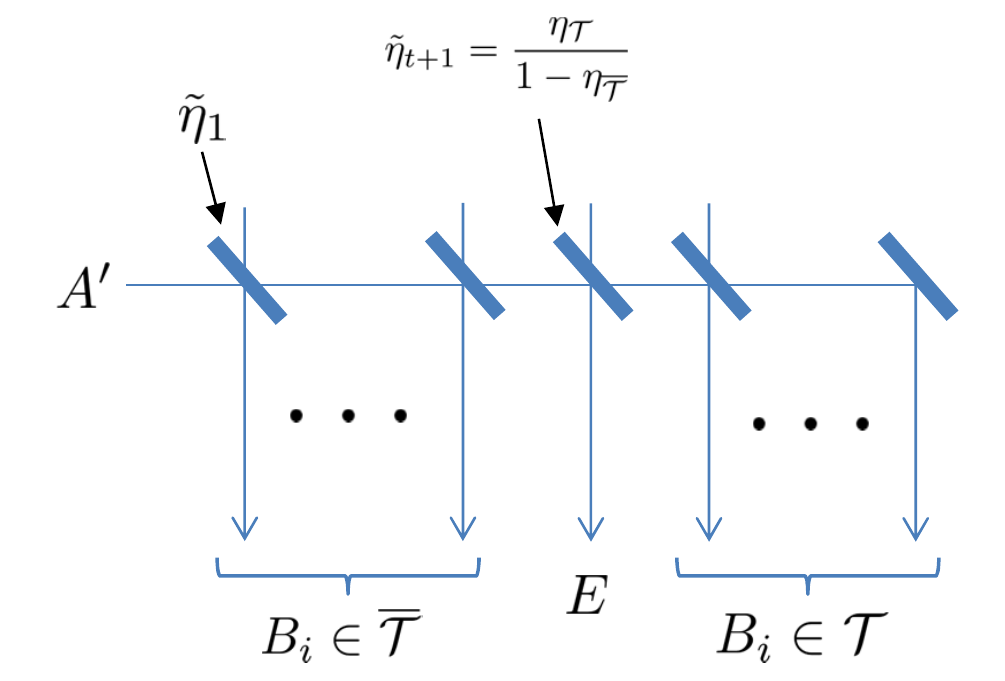}   %
\caption{\label{fig3}
Implementation of the 1-to-$m$ pure-loss 
bosonic broadcast channel. 
}
\end{center}
\end{figure}
\end{proof}

\end{document}